\documentclass{PoS}

\usepackage{amsmath,amssymb,graphicx}

\newcommand{\bgam}{\beta_\gamma}
\newcommand{\meff}{m_\mathrm{eff}}
\newcommand{\fgam}{F_\gamma}

\title{Constraints on chameleons and axion-like particles from the GammeV experiment}

\ShortTitle{The GammeV Experiment}

\author{\speaker{Jason H. Steffen}\\
        Fermilab Center for Particle Astrophysics\\
        E-mail: \email{jsteffenATfnalDOTgov}}

\author{For the GammeV Collaboration}

\abstract{
We present the most recent results of both aspects of the GammeV experiment.  The first aspect is a search for axion-like particles using a variable baseline, "light-shining-through-a-wall" technique.  This search excludes the particle interpretation of the PVLAS signal with high confidence.  The second aspect of the GammeV experiment is a search for chameleon particles, scalar particles which may be responsible for the dark energy of the universe.  This is accomplished by looking for a characteristic afterglow signature from a "particle-in-a-jar" experiment whereby chameleon particles become trapped in a region with a high magnetic field and slowly decay into detectable photons.  This is the first use of this experimental technique to probe for these particles.  We place interesting limits on a limited range of general chameleon models.  These limits are complimentary to other experiments, such as torsion pendulum experiments, which probe for forces that would result from new scalar particles.
}

\FullConference{Identification of dark matter 2008\\
		 August 18-22, 2008\\
		 Stockholm, Sweden}

\begin{document}

\section{Introduction}

The recent PVLAS experiment reported a signal in a photon oscillation experiment \cite{Zavattini:2005tm} that was consistent with the presence of a new spin-0 particle \cite{Raffelt:1987im}.  While PVLAS was not able to determine the parity of the new particle, their four measurements gave a consistent picture of a low-mass scalar $m_\phi\sim 1.2\mbox{ meV}$ with a rather large two-photon coupling $g \sim 2.5\times 10^{-6} \mbox{ GeV}^{-1}$.  Several experiments, including the GammeV experiment, set out to test the particle interpretation of the PVLAS signal \cite{Patras}.  These experiments make use of the Primakoff effect by which a photon may oscillate into an axion, which passes through an optical barrier and oscillates back into a photon on the far side (i.e. a ``light shining through a wall'' or LSW experiment).  Here we report the results of the GammeV experimental probe for such an axion-like particle (ALP) which used the LSW principle as its fundamental design consideration.  We also report the results for the GammeV search for chameleon particles.

Chameleon particles are scalar particles with a matter coupling and a nonlinear self interaction which give the field an environment-dependent effective mass.  This dependence on the environment allows the chameleon particles to have a small mass in intergalactic space, but large masses in generic terrestrial experiments.  Thus, chameleons may play an important role cosmologically~\cite{chamKW1,chamKW2} yet remain unobserved in the laboratory due to their large effective mass inside terrestrial experiments~\cite{chamKW1,chamKW2,chamcos,UpadhyeGubserKhoury,Adelberger2007}.  For most chameleon models, the effective mass of the chameleon scales as the local energy density to some power, $m_{\text{eff}} \sim \rho^{\alpha}$ where $\alpha$ is typically of order unity.  Generally, chameleons also couple to photons via terms such as $\phi F^{\mu\nu} F_{\mu\nu}$ and $\phi {\tilde F}^{\mu\nu}F_{\mu\nu}$, for scalars and pseudoscalars, respectively; here, $F_{\mu\nu}$ is the electromagnetic field strength tensor and ${\tilde F}_{\mu\nu}$ its dual.

This electromagnetic coupling allows photons to oscillate into chameleons and back in the presence of an external magnetic field.  As the chameleon penetrates the material of the apparatus, such as the walls and windows of a vacuum chamber, their mass grows sharply and they reflect.  Thus, a chameleon with an energy less than the effective mass in a material will be completely reflected by that material, allowing them to be trapped inside a ``jar''.  Chameleons produced in the jar from photon oscillation will be confined until they regenerate photons, which emerge as an afterglow once the original photon source is turned off~\cite{Ahlers:2007st,Gies:2007su}.

\section{GammeV Apparatus}

\begin{figure}[t]
\includegraphics[width=0.45\textwidth]{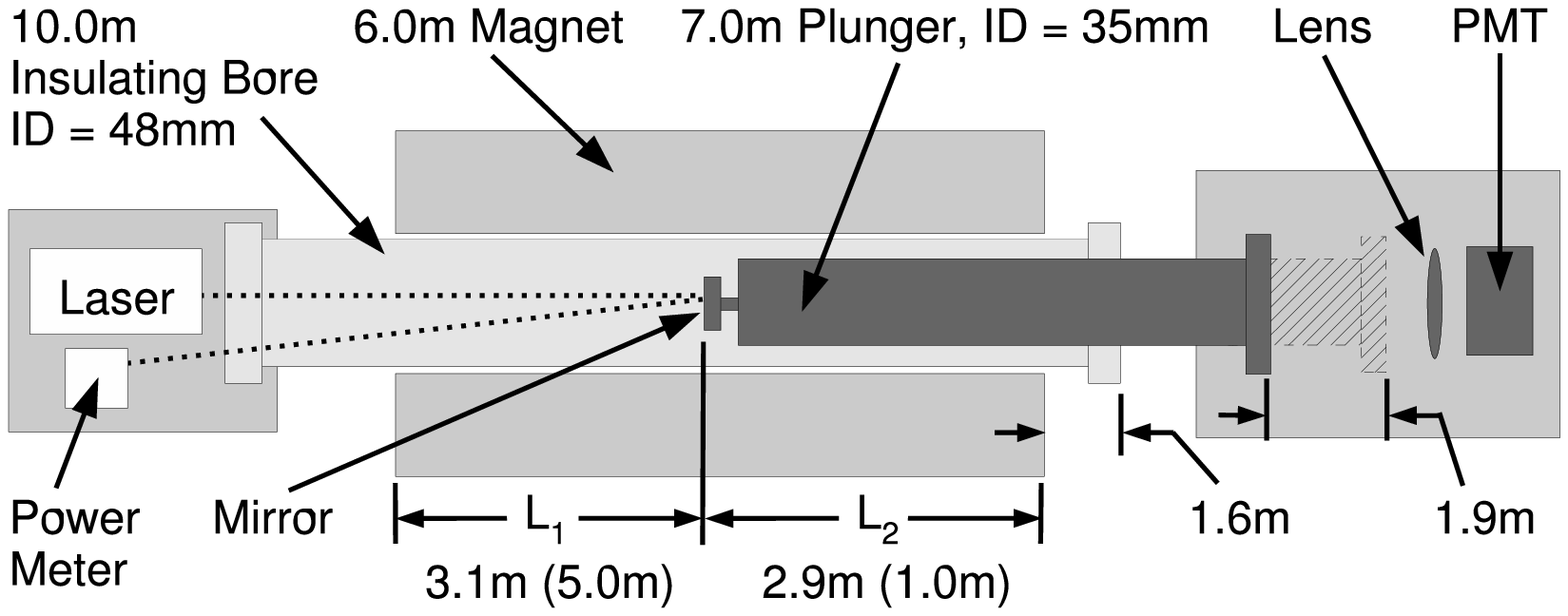}
\qquad
\qquad
\includegraphics[width=0.45\textwidth]{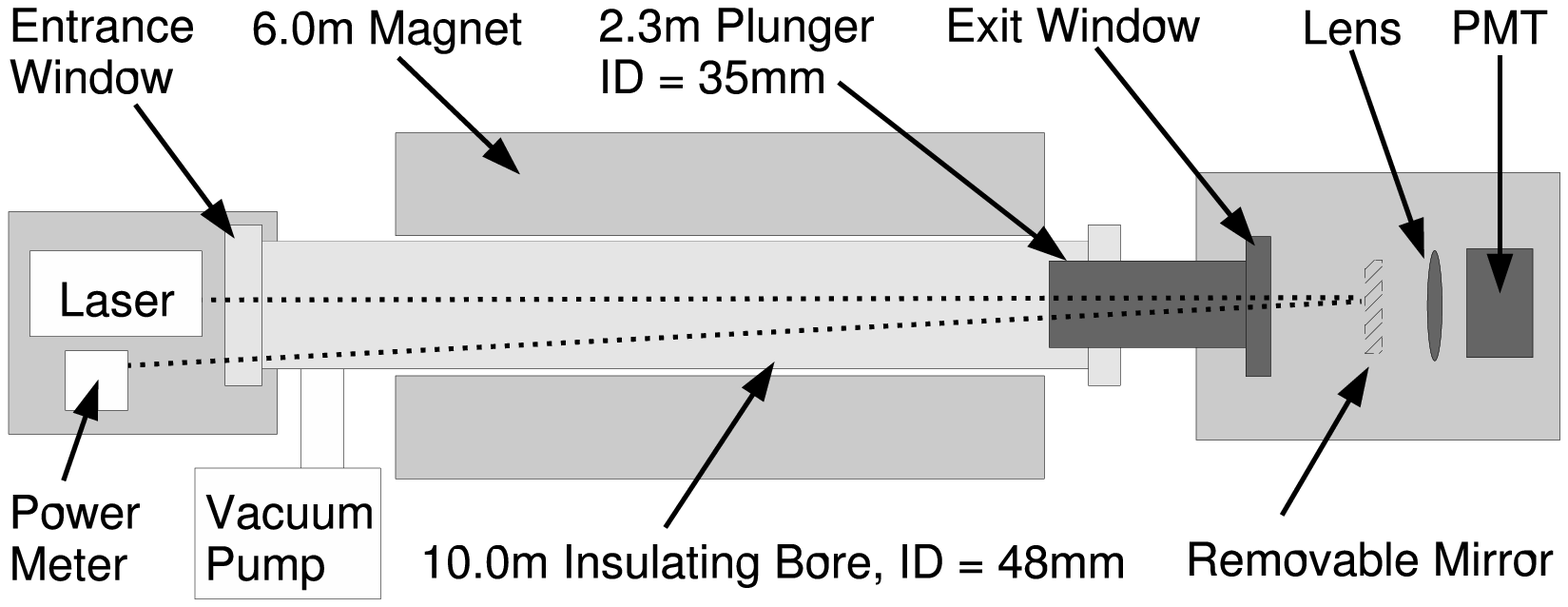}
\caption{Diagram (not to scale) of the experimental apparatus.  (Left) The LSW experimental setup.  The plunger can be translated roughly 2 meters in order to modify the magnetic field lengths.  (Right) The chameleon setup.  The short plunger has an open aperture which allows the afterglow photons to escape towards the PMT.\label{cartoons}}
\end{figure}

The GammeV apparatus, described in \cite{Chou:2007zzc}, for both the LSW configuration and the chameleon configuration is shown in Figure~\ref{cartoons}.  It consists of a long cylindrical vacuum chamber inserted into the bore of a $B=5\mbox{ T}$, $L=6\mbox{ m}$ Tevatron dipole magnet.  The entrance and exit of the chamber are sealed with BK7 glass vacuum windows.  We use a $20$~Hz pulsed Nd:YAG laser that emits light of wavelength $532$~nm into the chamber at a rate of $\fgam \sim 10^{19}$~photons/sec.  

The novel features of the GammeV LSW design is the moveable plunger and the pulsed laser.  Since the photon-ALP conversion is an oscillation phenomenon, a fixed magnet length will be insensitive to certain masses of the ALP---a complete oscillation cycle produces a state with zero ALP amplitude.  The plunger design allows us to change the lengths of the magnetic field regions so that there are no gaps in our sensitivity.  The pulsed laser is effective at removing spurious, background PMT hits because the arrival of the ALP signal should be coincident with the arrival of the $\sim 10$ns-wide laser pulses.

For the chameleon experiment, the salient feature is the mass density within the vacuum chamber.  Our sensitivity to a given chameleon model depends upon how much the chameleon mass rises when traveling between the lowest density region within the vacuum chamber, where the chameleons have the smallest effective mass, and the lowest density region outside the vacuum chamber, the boundary where chameleons would be most likely to escape.  These considerations limit our chameleon sensitivity to models where $\alpha \gtrsim 0.8$ ($\alpha$ being defined in the introduction).

\section{Results for Axion-like Particles}

For the LSW experiment, we gathered 20 hours of data in each of four configurations (two polarizations for scalar and pseudoscalar couplings, and two plunger positions).  We count the PMT triggers within the 10 ns coincidence time window, defined {\it a priori} by the our calibration data.  The expected background is measured using the dark counts in time bins outside of the coincidence window.  The data, summarized in Table~\ref{axiontab}, show no excess counts above background in any of the four configurations.

\begin{figure}[t]
\includegraphics[width=0.48\textwidth]{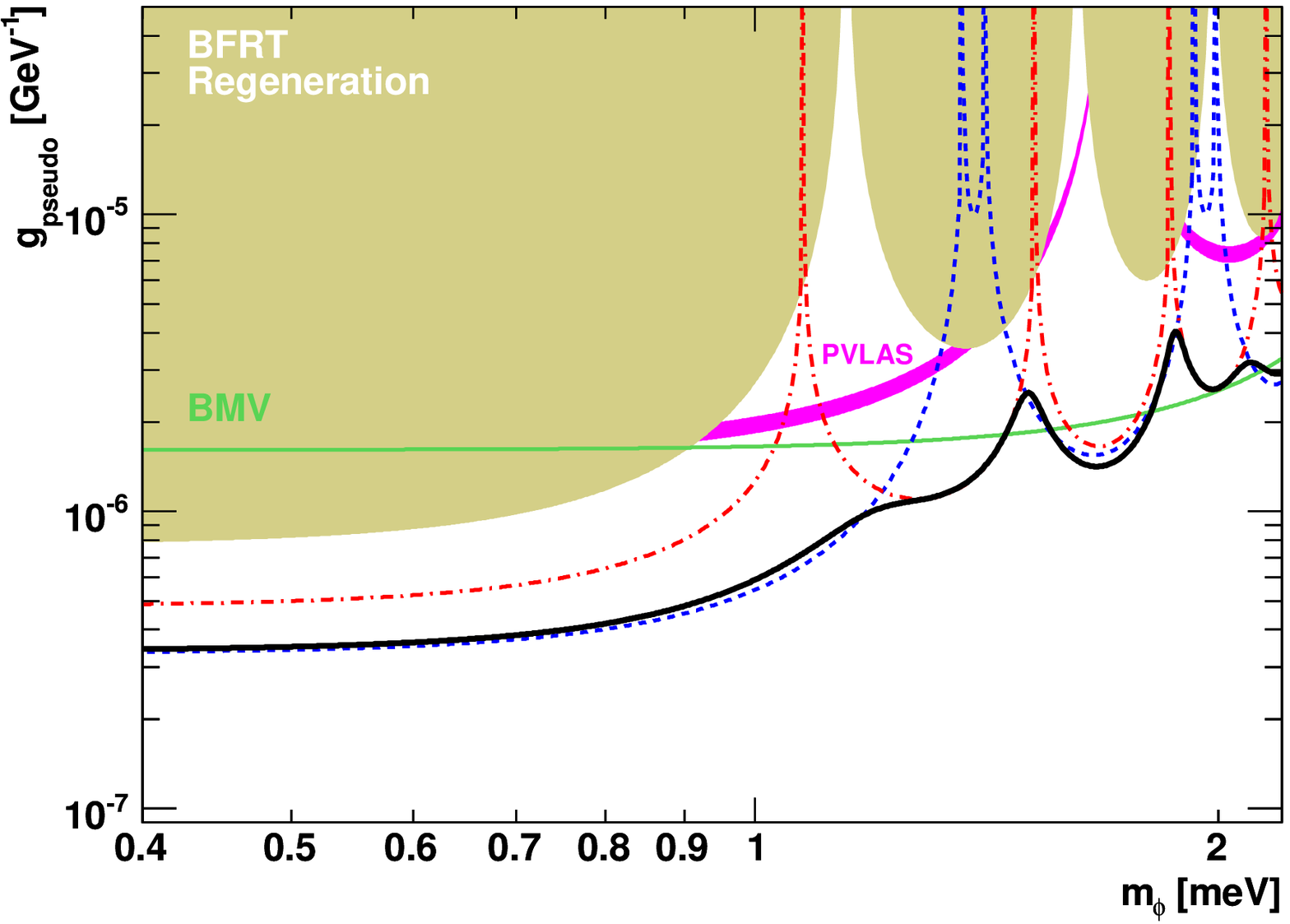}
\includegraphics[width=0.48\textwidth]{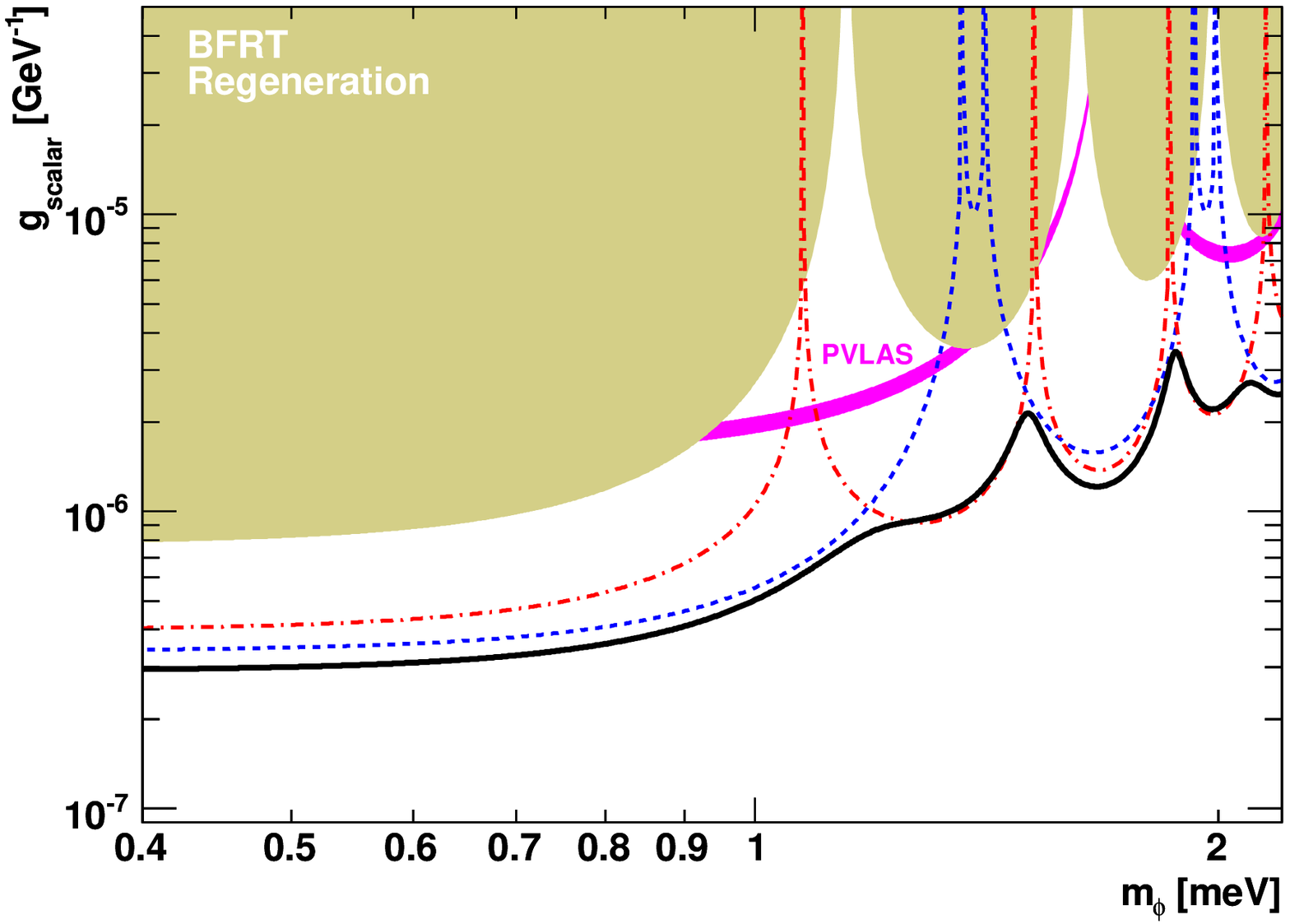}
\caption{3$\sigma$ upper limits on the couplings of pseudoscalar (left) and scalar (right) axion-like particles as a function of the particle mass from the GammeV experiment.\label{axionlims}}
\end{figure}

We use the Rolke-Lopez method \cite{Rolke:2004mj} to obtain limits on the regeneration probabilities and obtain the corresponding $3\sigma$ upper bounds on the coupling $g$ as a function of the ALP mass, $m_\phi$.  The relative systematic uncertainties of $12\%$ on the photon transport and detection, and $3\%$ on the laser power measurement are incorporated in the limits.  The GammeV limits are shown in Figure \ref{axionlims} along with the PVLAS 3$\sigma$ signal region, the BFRT 3$\sigma$ regeneration limits, and the $99.9\%$ limit on pseudoscalar couplings from BMV.  As expected, the regions of insensitivity for one plunger position are well-covered by using the other plunger position.  Data from both plunger positions are combined and analyzed jointly to produce the combined limit curve.  The weakly-interacting axion-like particle interpretation of the PVLAS data is excluded at more than $5\sigma$ by GammeV data for both scalar and pseudoscalar particles.

\begin{table}
\caption{\label{axiontab} Summary of data in each of the 4 configurations for the ALP search.  The stated coupling limits are for asymptotically small ALP masses.}
\begin{center}
\begin{tabular}{|l|c|c|c|c|}
\hline
Configuration&$\#$ photons&Est.Bkgd&Candidates&g[GeV$^{-1}$]\\
\hline
Horiz.,center&$6.3\times 10^{23}$&$1.6$&1&$3.4\times 10^{-7}$\\
Horiz.,edge&$6.4\times 10^{23}$&$1.7$&0&$4.0\times 10^{-7}$\\
Vert.,center&$6.6\times 10^{23}$&$1.6$&1&$3.3\times 10^{-7}$\\
Vert.,edge&$7.1\times 10^{23}$&$1.5$&2&$4.8\times 10^{-7}$\\
\hline
\end{tabular}
\end{center}
\end{table}

\section{Results for Chameleon Particles}

For the chameleon experiment we populate the vacuum chamber ``jar'' with chameleons by continuously shining the laser into the vacuum chamber for $\sim 5\mbox{ h}$.  After emerging through the exit window of the chamber, the beam is reflected back through the chamber in order to increase the chameleon production efficiency and facilitate monitoring of the laser power.  During the afterglow phase of the experiment, the laser is turned off and the low noise PMT placed at the exit window is uncovered.  Chameleons interacting with the magnetic field oscillate back into photons, a fraction of which can be detected by the PMT.

Data are taken in two separate runs, one for each laser polarization, to search for pseudoscalar and scalar chameleons.  Table \ref{chamtab} shows relevant data for both runs (as reported in~\cite{Chou:cham} including: the total integration time during the filling stage, the total number of photons which passed through the chamber, a limit on the vacuum quality (which can affect the coherence length of the photon-chameleon oscillations), the length of the afterglow observation run, the time gap between filling the chamber and observing the afterglow, the mean afterglow rate, and the limits on $\bgam$, the chameleon to photon coupling for coherent oscillations.

\begin{figure}[t]
\includegraphics[width=0.48\textwidth,angle=270]{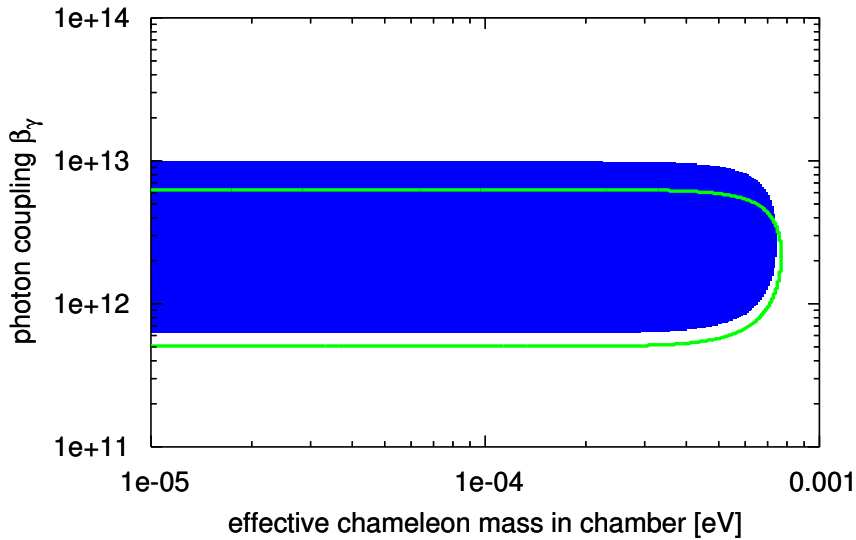}
\caption{Region excluded by GammeV to $3\sigma$ for pseudoscalar particles (solid blue region) and for scalar particles (region between green lines).  Chameleon interactions with the chamber walls typically prevent $\meff$ from dropping below $\sim 1/R \sim 10^{-5}$~eV. \label{chamresults}}
\end{figure}

The dominant uncertainty in our measurements of the chameleon afterglow rate is the systematic uncertainty in the PMT dark rate, $12$Hz.  Thus our $3\sigma$ upper bound on the mean afterglow rate is 36 Hz above the mean of the data for each run, after the background dark rate has been subtracted.  For each $\meff$ and $\bgam$ we compute the total number of excess photons predicted within the observation window and compare that afterglow signal to the mean signal observed by the PMT.  Figure~\ref{chamresults} shows the regions excluded by GammeV in the $(\meff,\bgam)$ parameter space for scalar and pseudoscalar chameleon particles.  For $\bgam \gtrsim 10^{13}$ our sensitivity diminishes because the chameleon decay time in GammeV could be less than the few hundred seconds required to switch on the PMT.  These are the first experimental results to specifically probe for chameleon particles and the first search for such light scalars using the afterglow phenomenon.

\begin{table}[t]
\caption{\label{chamtab} Summary of data for both configurations of the chameleon experiment.}
\begin{center}
{\footnotesize
\begin{tabular}{|l|c|c|c|c|c|c|c|}
\hline
Config. & Fill (s) & $\#$ photons & Vac. (Torr) & Data (s) & Offset (s) & Rate (Hz) & g[GeV$^{-1}$] \\ \hline
Pseudo  & 18300         & $2.0e23$ & $<2e$-7 & 3602 &  319 & 123 & $6.2e11 < \bgam < 1.0e13$ \\
Scalar        & 19100         & $2.0e23$ & $<1e$-7 & 3616 & 1006 & 101 & $5.0e11 < \bgam < 6.2e12$ \\ \hline
\end{tabular}
}
\end{center}
\end{table}

\section{Prospects}

With the initial GammeV experiment winding down, we are looking at new searches for axions or axion-like particles.  These new experiments will build upon the experience of the first GammeV suite.  The next generation search for axions which will use coupled, Fabry-Perot cavities to resonantly enhance the axion production and regeneration; this design will improve the sensitivity of the experiment by several orders-of-magnitude, possibly making it more sensitive than astrophysical and solar constraints on axions~\cite{vanbibber}.  The enhanced chameleon experiment, with modest changes to the existing apparatus, will also improve upon our current limits.  More importantly, this new chameleon experiment will be sensitive to a wider range of chameleon models including cosmologically interesting, chameleon dark energy models as well as several classes of generic scalar field models that have not been tested by previous experiments.


\begin{thebibliography}{99}

\bibitem{Zavattini:2005tm}
  E.~Zavattini {\it et al.}  [PVLAS Collaboration],
  Phys.\ Rev.\ Lett.\  {\bf 96}, 110406 (2006)
  [arXiv:hep-ex/0507107].

\bibitem{Raffelt:1987im}
  G.~Raffelt and L.~Stodolsky,
  Phys.\ Rev.\  D {\bf 37}, 1237 (1988).

\bibitem{Patras}
   LIPSS, OSQAR, ALPS, BMV presentations at the 3rd Joint ILIAS-CERN-DESY Axion-WIMPS Training Workshop, University of Patras, Greece, 19-25 June 2007.

\bibitem{chamKW1}
J. Khoury and A. Weltman,
  Phys. Rev. Lett. \textbf{93} 171104 (2004).

\bibitem{chamKW2}
J. Khoury and A. Weltman,
  Phys. Rev. D \textbf{69} 044026 (2004).

\bibitem{chamcos}
P. Brax, C. van~de Bruck, A.-C. Davis, J. Khoury, and A. Weltman,
  Phys. Rev. D \textbf{70} 123518 (2004).

\bibitem{UpadhyeGubserKhoury}
A. Upadhye, S.~S. Gubser, and J. Khoury,
  Phys.~Rev.~D \textbf{74} 104024 (2006).

\bibitem{Adelberger2007}
E.~G. Adelberger \textit{et~al.}
Phys.~Rev.~Lett. \textbf{98} 131104 (2007)
  [arXiv:hep-ph/0611223].

\bibitem{Ahlers:2007st}
M. Ahlers, A. Lindner, A. Ringwald, L. Schrempp, and C. Weniger,
  Phys.~Rev.~D. \textbf{77} 015018 (2008) [arXiv:0710.1555 [hep-ph]].

\bibitem{Gies:2007su}
H. Gies, D.~F. Mota, and D.~J. Shaw,
  Phys.~Rev.~D. \textbf{77} 025016 (2008) [arXiv:0710.1556 [hep-ph]].

\bibitem{Chou:2007zzc}
A.~S. Chou and \textit{et. al}
  Phys.~Rev.~Lett. \textbf{100} 080402 (2008) [arXiv:0710.3783 [hep-ex]].

\bibitem{Rolke:2004mj}
  W.~A. Rolke, A.~M. Lopez and J. Conrad,
  Nucl.\ Instrum.\ Meth.\  A {\bf 551}, 493 (2005)
  [arXiv:physics/0403059].

\bibitem{Chou:cham}
A.~S. Chou and \textit{et. al}
  Phys.~Rev.~Lett. Submitted (2008) [arXiv:0806.2438 [hep-ex]].

\bibitem{vanbibber}
 Sikivie, P., Tanner, D. B., and van Bibber, Karl,
 Phys.~Rev.~Lett, \textbf{98} 172002 (2007).



\end{thebibliography}
\end{document}